\documentclass[a4paper,fleqn,usenatbib]{mnras}
\usepackage{newtxtext,newtxmath}
\usepackage{hyperref}
\usepackage{epsfig}
\usepackage{subfigure}
\usepackage[T1]{fontenc}
\usepackage{xcolor}
\usepackage{graphicx}	
\usepackage{amsmath}	
\usepackage{amssymb}	
\usepackage{breakurl}   

\hypersetup{pdftitle=Temporal Variabilities of XTE J1650-500}

\title[Temporal Variabilities of XTE J1650-500]
{Time-domain variability properties of XTE J1650-500 during its 2001 
outburst: Evidence of disc-jet connection }

\author[Chatterjee, Dutta, Nandi \& Chakrabarti]{
Arka Chatterjee,$^{1}$\thanks{\href{mailto:arkachatterjee@bose.res.in}{arkachatterjee@bose.res.in}}
Broja G. Dutta,$^{2,3}$\thanks{\href{mailto: brojadutta@gmail.com}{brojadutta@gmail.com}}
Prantik Nandi,$^{1}$
and Sandip K. Chakrabarti $^{3}$\\
$^{1}$Department of Astrophysics \& Cosmology, S. N. Bose National Centre For Basic Sciences, Kolkata, 700106, India.\\
$^{2}$Department of Physics, Rishi Bankim Chandra College, Naihati, W.B, 743165, India.\\
$^{3}$Indian Centre for Space Physics, Kolkata, 700084, India.
}

\date{Accepted XXX. Received YYY; in original form ZZZ}
\pubyear{2020}
\begin{document}
\label{firstpage}
\pagerange{\pageref{firstpage}--\pageref{lastpage}}
\maketitle

\begin{abstract}
Evolution of variability properties of Galactic transient sources is 
a diagnostic tool to understand various regimes of the accretion flow 
and its dynamics close to the central black hole. In this paper, we 
concentrate on the variability properties of the X-ray transient XTE 
J1650-500 and investigate the evolution of viscous delay, time lag, 
QPO frequency, and their energy dependence throughout the rising phase 
as observed by {\it RXTE} during its outburst in 2001. Our analysis reveals (1) a delay of
$12\pm1$ days between one day averaged hard (5-12 keV) and soft 
(1.5-3 keV) photon light-curves as observed by {\it RXTE}/ASM; (2) 
QPOs with high rms values are observed in lower energy (4-8 keV) range; 
(3)  the QPO frequencies and associated time lags were anti-correlated 
during the initial days of the rising phase, and later on, they were 
found to be correlated; (4) the time lags of iron line photons with
respect to hard and soft photons remained almost constant during the 
initial days of hard state and the lag magnitude increased during the 
state transition. We perform comparative studies with outbursts of 
GX 339-4 and XTE J1550-564. We find the evolution of time lags associated 
with the QPO characteristics during the outburst -- stronger QPOs at low energy, 
and constant lags of broad Fe-line photons present a unique nature of 
outburst profile in XTE J1650-500. The possible cause of such variabilities 
is explained by considering disc-jet geometry closer to the central black hole.
	
\end{abstract}


\section{Introduction}
\label{intro}
Variability properties of Galactic X-ray transients (XRT) in time-scales 
of milliseconds to days are extensively reported in the literature 
\citep{Mi88, van87, sm02}. Earlier, it was assumed that the instabilities in the 
standard accretion disc might generate the observed variability \citep{SS73, LE74}. 
However, proper cause of variabilities remained uncertain 
due to possible non-linear mechanisms of the physical processes which may occur during accretion
onto compact objects. This variability study includes the study of Power Density 
Spectrum (hereafter PDS), the time/phase lag spectrum, and their dependence on the 
energy of the emergent photons from a region closer to the central engine. From the PDS, 
one can study the Quasi-Periodic-Oscillations (hereafter QPOs) whose frequencies range 
from a few mHz to $\sim$ a few hundred Hz \citep{BH90, van04, MR06} and their dependence on 
the energy \citep{Be97, CZCM99, Ru99, Ka15} of the emergent radiation.

The nature of the evolution of the C-type LFQPOs during the Low Hard State (hereafter LHS) 
to Intermediate State (IS) at the onset and decay phase of an outburst for several
Galactic black hole transients, such as GRO J1655-40, XTE J1550-564, 
GX 339-4, H1743-322, and persistent source like GRS 1915+105 is 
very generic irrespective of the mass and inclination and 
follow a well-established pattern (see, \cite{Mo15} and 
references therein). The origin of such LFQPOs remained enigmatic, and 
several models are proposed to address this issue. Among them, the epicyclic motion 
of dense matter around the black hole or Neutron stars \citep{SV99a, in09} stands 
out as a relativistic effect. The formation of LFQPOs due to the coupling of radiation
with the motion of astrophysical fluids around compact objects are also in the 
literature. 
In that context, Two Component Advective Flow (TCAF) model proposed by \cite{CT95} 
can be employed to understand the LFQPOs and their evolution where the Compton cloud is 
 formed at the CENtrifugal pressure supported BOundary Layer 
or CENBOL, (see \cite{Ch99}; \cite{cm00}) and at least C-type QPOs are produced by the resonance oscillation 
due to a rough agreement between the compressional heating and radiative cooling of this 
Compton cloud as shown by numerical simulations \citep{Mo94,ga14}. 
In TCAF paradigm, the evolution of the QPO frequency was explained by steady
radial drifting of the shock front caused by changes in post-shock cooling
as the mass accretion rates evolve. From \cite{cm00, vad01, rao00, ch05}, it was 
established that the Comptonized photons produce QPOs, and thus they are 
intrinsically related to Compton cloud size. \cite{he15, Mo15} reported the dependence 
of spectra and the amplitude of the QPOs over the inclination of the source and 
\citet{s06, iv15} strongly suggested a geometric origin of QPOs. 
\cite{tp99, c10} suggested a connection between the LFQPOs and 
the variation of luminosity due to the change in the mass accretion rate. 
Under these circumstances, obtaining a clear picture of the lags associated 
with the LFQPOs becomes even more critical as it could constrain the geometry 
of the accretion disc-Compton cloud-jet system. 

Time/Phase lags are calculated using cross Fourier spectrum of X-ray light curves of 
different energy bands, and it is the difference in time of arrival between soft and 
hard photons \citep{Mi88}. Hard lag or positive lag implies hard photons arrive 
later than their softer counterparts, and the opposite is true for soft or negative 
lag. For Galactic XRTs, the Keplerian disc flux dominates in the $0.1-5.0$ keV energy 
band and harder photons produced via the inverse Compton process \citep{ST80} 
dominate in the energy band above five keV. The time lags were first explained to be 
due to the Comptonization of soft
seed photons by hot electrons, known as `Compton reverberation' \citep{p80, Mi88} which
naturally produces hard time lags. Several models are proposed \citep{CCZ99, NWD99, PF99} 
to explain the hard and soft lags associated with QPOs observed in Galactic binary systems. 
Propagating perturbation model suggested by \cite{BL99, l00} refers to a change in the
phase lag sign with the QPO frequency triggered by a propagation direction 
reversal when the QPO frequency is close to the crossover frequency. 

Contrary to QPO evolution which is monotonic in rising or declining phases, 
time lag evolution starts with maximum value and decreases often switching 
the sign of the lag. They are observed in black hole transients such as 
XTE J550-564, GX 339-4, H1743-322 and persistent sources like GRS 1915+105  
\citep{dc16, dpc18}. A generic feature was observed during the rising 
state of the outburst where time lag started from a maximum value when first QPO was 
detected, i.e., at the minimum QPO frequency and smoothly decreased to a 
minimum as the QPO frequency increases to the maximum. Through Monte-Carlo 
simulations, considering the Comptonization, gravitational bending, 
and reflection, \cite{Ch17b} found the anti-correlations between time lag and 
QPO frequency as a result of the reducing size of the Compton cloud. 
The opposite feature was observed during the declining state, where time lag starts 
from a minimum and rises to a maximum as the QPO frequency shows the opposite behaviour. 

\subsection{XTE J1650-500}
\label{intro:source}
In the present paper, we study the time/phase lag behaviour during 
the {\it only} outburst of XTE J1650-500, which took place in 2001. 
The Galactic X-ray transient, XTE J1650-500 
was discovered by the All-Sky Monitor (ASM) on the {\it RXTE (Rossi X-ray 
Timing Explorer)} satellite on 2001 September 5 \citep{Re01}, and it 
went through all the usual black hole X-ray spectral states \citep{MR06} before 
returning to the quiescent state in June 2002 \citep{Ro04}. 
During this outburst, {\it RXTE} observed the source continuously, {\it BeppoSAX}
observed it three times, and {\it XMM-Newton} observed it once.
\cite{Sa02} reported that the mass is greater than $3 M_{\odot}$ 
and inclination is less than $40^{\circ}$. Later, \cite{Or04} 
reported the orbital inclination to be $\sim 50^{\circ} \pm 3^{\circ}$, 
and the upper limit of mass of the central black hole is estimated 
to be $\sim 7.3 M_{\odot}$. However, \cite{Or04} 
expressed their concern and suggested the possibility of the source to have much 
less mass if the optical emission during the quiescent state is dominated 
by the accretion disc rather than the companion itself. Using {\it XMM-Newton} data 
\cite{Mi02} observed a broad iron $K{\alpha}$ emission line and concluded 
that the compact star could be an extreme Kerr black hole with a spin parameter 
$a\sim 0.998$. Similar conclusions were drawn from {\it BeppoSAX} data by 
\cite{Mi04}, where the effects of light bending are discussed as well. 
\cite{Ho03} reported high-frequency quasi-periodic oscillations (HFQPOs) 
along with sub-harmonics and higher-harmonics. \cite{To04} identified a short time-scale 
($\sim 100$s) X-ray flares and long time-scale oscillations during the declining phase. 
\cite{Ka03} found the low-frequency QPOs in the range of $4-9$ Hz during the Low Hard 
State (LHS) of outburst decay phase and suggested a possible strong disc-jet connection in this 
object.  

\begin{figure*}
\begin{center}
\includegraphics[height=1.2\columnwidth,width=2.0\columnwidth]{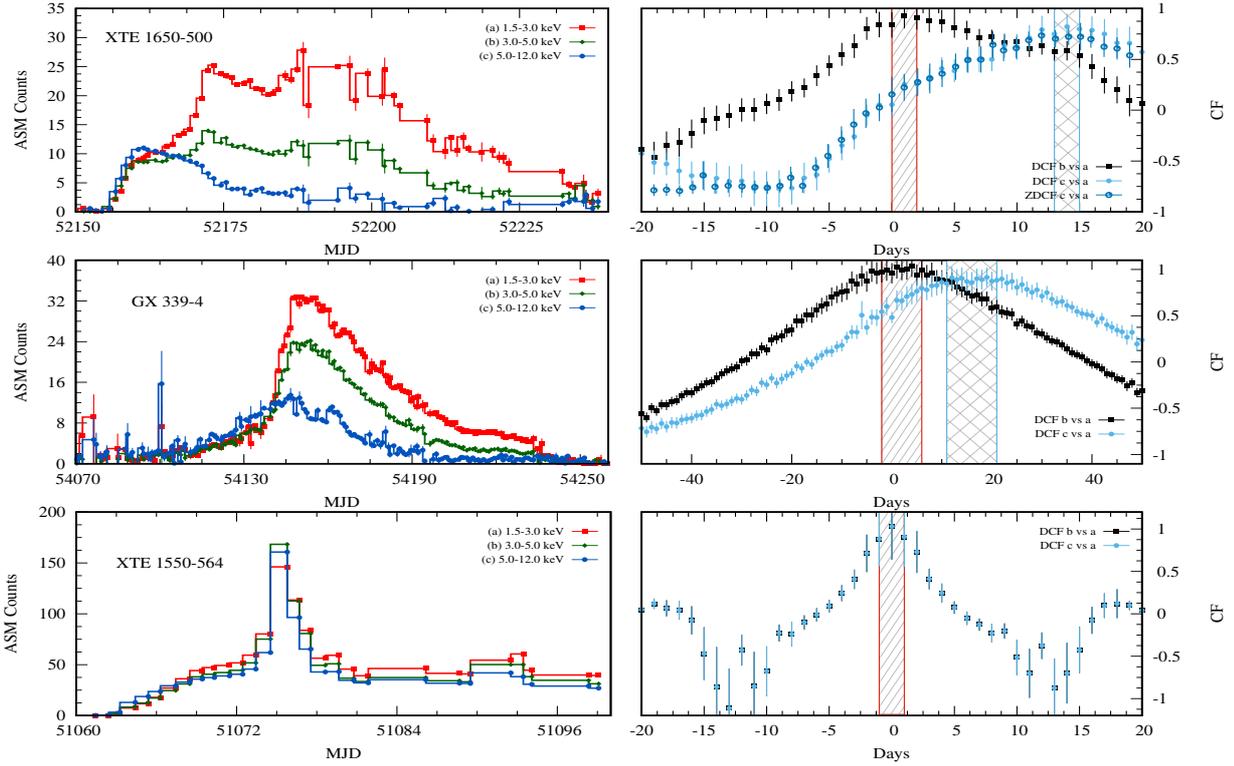}
\caption{(Left Panels) Comparison of ASM light curves at 
three different energy bands (a: 1.5-3.0 keV, b: 3.0-5.0 keV \& c: 5.0-12.0 keV) of three 
outbursting sources. (Right Panels) Corresponding cross-correlation function (CF)
between a and b bands (black) and a and c bands (sky-blue) are plotted. Shaded regions 
represent the error bar. 
\label{fig1}}
\end{center}
\end{figure*}

Variability in the radio frequency domain found during the 2001 outburst 
was reported in \cite{Co04}, where jet emission is also seen in thermal 
disc dominated state. \cite{Ro05} investigated the iron line flux to 
diagnose the effect of the light-bending scenario in the evolution of 
XTE J1650-500. Using the BeppoSAX data, \cite{Mo09} found that while the 
source had undergone a transition from the hard state to softer states, the 
power spectra as a whole drifted towards the higher frequency regime and 
this behaviour was interpreted as the reduction in the radius of the Compton 
cloud. More quantitative values on the reduction of Compton cloud ($\sim$ 23 
times of the initial size) was presented in \cite{YW12}.

\cite{DG06} observed the iron line having an equivalent width of 
$190^{+30}_{-20}$ eV using {\it BeppoSAX} data.
In their work, the best fit $\chi^2$ was observed at an inclination
less than 30\textdegree ~ where relativistic smearing was considered. 
However, considering outflowing warm absorber, the fit becomes insensitive with 
the inclination and fits well in the 27\textdegree-70\textdegree ~ domain. The
degeneracy in the inclination angle, originated mostly from spectral fitting, 
suggests an exploration of all possible origins of broad iron lines; (a) enhanced 
reflection caused by gravitational bending \citep{FV03}, (b) jet Comptonization
mechanism \citep{LT07}. It was also observed that the disc-jet coupling 
affects the timing properties \citep{Ga08, AM15, Ve17, Re18, pa19} as well as 
the spectral properties \citep{Me03, Co03}. In the case of 
XTE J1650-500, \cite{Cu12} reported that the jet reprocessed the X-rays 
emitted from the Compton cloud and the contribution of jet reprocessing 
can be seen in the optical-NIR flux during the rising phase. It is thus clear 
that a consensus regarding the emission properties of this source is still 
missing. Therefore, we examined the rising phase of 2001 outburst, 
when the maximum contribution from the jet can be seen in the X-ray flux, 
through the timing properties keeping a combined disc-jet-Compton 
cloud paradigm in mind. 

In this article, we analysed the variability properties such as the time delay, 
QPOs and phase/time lag during its low/hard state to intermediate states 
of 2001 outburst. In \S \ref{results}, we outline the time
domains, e.g., long-time and for short-time-scales. In \S \ref{time_delay}, we 
present the results of ASM data analysis and compare other outbursts with our
current studies. Results obtained from short-time scale variabilities are presented 
in \S \ref{qpo_power_lag}.  Later, in \S \ref{discussion}, we discuss the 
possibility of viscous time-scale causing the delay observed from one day 
averaged light-curve. We examine the combined effects of the Comptonizing 
region and outflows in the evolution of QPOs and as well as 
the associated time lags. Finally, we draw our conclusions in \S \ref{conclusion}.
\section{Time-domain Analysis}
\label{results}

We divided the timing analysis into two parts. First, we analysed the ASM data and performed
the cross-correlation analysis. Later, we examined each observation, PDS, {\it rms}, 
and time lags in various energy bands. 

\subsection{Long-time-scale}
\label{time_delay}
We performed discrete cross-correlation ({\tt DCF}, \cite{ek88}) between the energy bands 
of All-Sky Monitor (ASM)\footnote{\url{http://xte.mit.edu/ASM\_lc.html}} data, and compare the  
outburst profile of XTE J1650-500 with other outbursts of black hole candidates such as
 GX 339-4 and XTE J1550-564. We used the interpolated cross-correlation 
({\tt ICF}, \cite{ga87}) and $\zeta$-discrete cross-correlation function 
({\tt ZDCF}\footnote{{\tt ZDCF: }\burl{http://www.weizmann.ac.il/particle/tal/research-activities/software}}, 
\cite{A97}) for comparison. To remove the ambiguity of peak calculations in delay 
contributed by the skewness of the correlation patterns, we evaluated peak error 
(see, \cite{ga87} and references therein) 
\begin{equation}
\Delta_{p} = \frac{0.75\times\omega_c}{1+h(n-2)^{1/2}},
\end{equation}
\noindent where `$h$' refers to the peak value of the correlation, `$\omega_c$' is HWHM 
and `$n$' is the number of data points. We compared $\Delta_{p}$ with the bin width 
$\Delta_{ASM}$ ($\sim$ 1 day) of the ASM data. The larger of these two is considered 
as the delay error ($\Delta_{delay}$). Please see Table \ref{tab:delay} for details. 

\begin{table}
\centering

    \caption{Delay Estimation}     
    \label{tab:delay}

    \begin{small}
    \begin{tabular}{|c|c|c|c|c|c|c|}
    \hline
	    {\bfseries Sources}            & {\bfseries $\omega_c$}   & {\bfseries h}  & {\bfseries n }  & {\bfseries $\Delta_{p}$}  & {\bfseries $\Delta_{delay}$} &{\bfseries $\tau_{delay}$}\\
            {\bfseries }                     & {\bfseries days}       & {\bfseries }   & {\bfseries }    & {\bfseries days}          & {\bfseries days} & {\bfseries days}         \\
    \hline
	    \textcolor{black}{XTE J1650-500}       & 12.25            &   0.82         & 350             & 0.56                      &   1.0  & $0^{+1}_{-1}$     \\
	    \textcolor{cyan}{XTE J1650-500}        & 10.25            &   0.93         & 350             & 0.42                      &   1.0  & $12^{+1}_{-1}$    \\
    \hline
	    \textcolor{black}{GX 339-4}            & 24.25            &   0.91         & 180             & 1.38                      &   1.4   & $0^{+1.4}_{-1.4}$ \\
	    \textcolor{cyan}{GX 339-4}             & 20.65            &   1.00         & 180             & 1.08                      &   1.1   & $16^{+1.1}_{-1.1}$ \\
    \hline
	    \textcolor{black}{XTE J1550-564}       & 1.35             &  0.97          & 75              & 0.109                     &   1.0   &  $0^{+1}_{-1}$    \\
	    \textcolor{cyan}{XTE J1550-564}        & 1.35             &  1.00          & 75              & 0.106                     &   1.0   &  $0^{+1}_{-1}$    \\
    \hline
    \end{tabular}
    \end{small} 
\end{table}

\begin{figure*}
\begin{center}
\includegraphics[height=1.2\columnwidth,width=2.0\columnwidth,angle=0]{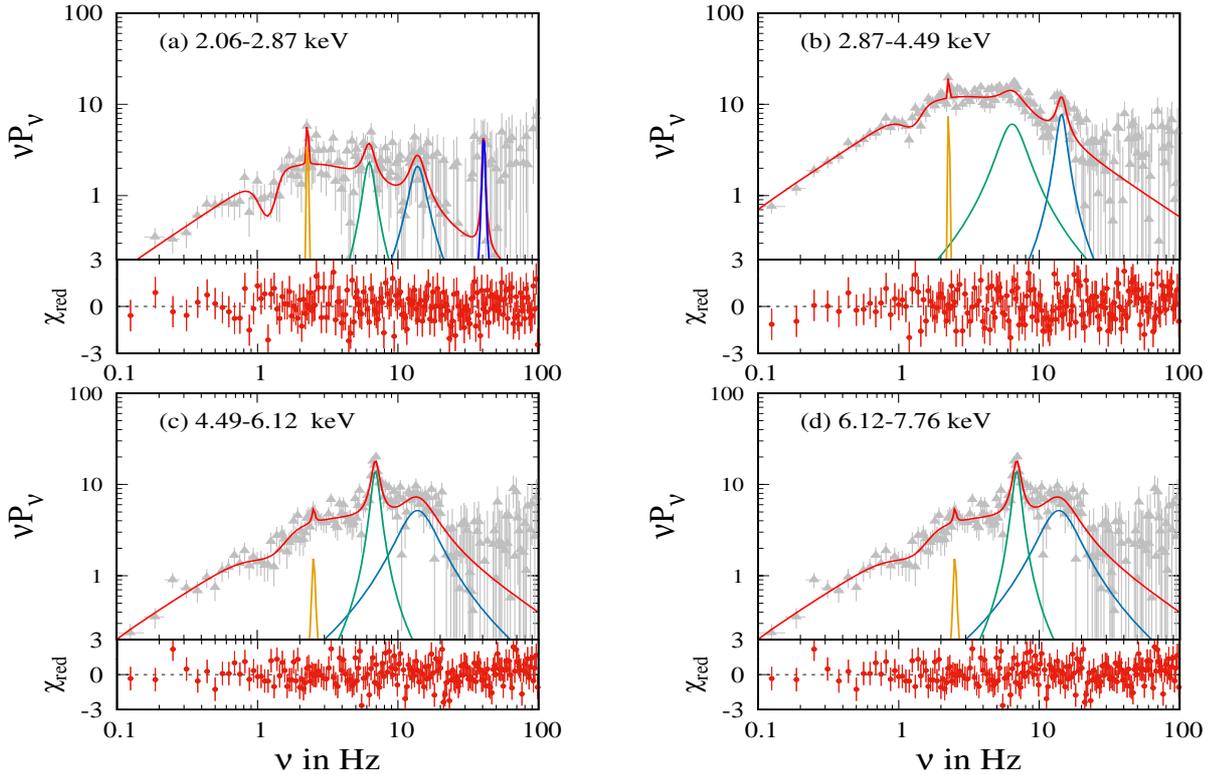}
\caption{Energy-dependent $\nu P(\nu)$ fitted with four Lorentzians. Corresponding 
observation ID 60113-01-13-02 has MJD 52171.7. Along with Lorentzians, it 
required multiplicative model {\tt gabs} at 1.19 Hz. Strength of the 
{\tt gabs} is reduced with increasing energy band.
\label{fig2}}
\end{center}
\end{figure*}

In the top left panel of Fig. \ref{fig1}, a clear distinction in progress 
between the harder (circle-blue) band and their softer counterpart (square-red) 
can be observed after MJD $\sim$52160. {\tt DCF} between a and b bands ($\rho$(a,b), 
square-black) presented in the top right curve shows no discernible time 
delay ($1\pm1$) between them as the zero-delay resides within the error bar. 
However, {\tt DCF} between a and c bands $\rho$(a,c) indicates a delay of $12\pm1$ 
days. It is to be noted that $\rho$(a,c) is also minimum around $ 9 $ days. 
To understand the delay sign, we performed both {\tt ICF} and {\tt ZDCF}. 
{\tt ZDCF} showed a plateau (blue-circle in the top-right panel) region 
beyond $ 9 $ days. Nevertheless, the positive $12\pm1$ days delay coincides 
with what was observed from DCF. Thus, a conclusive long-time-scale delay 
{\it sign} of this source can be predicted from the ASM studies, and there indeed 
exists a long-time-scale lag between the hard and the soft photons.

Unlike XTE J1650-500, GX 339-4 exhibited several outbursts during the 
operational period of RXTE. We performed {\tt DCF} on the ASM light curves 
for each outburst. No significant long time-scale delay was found in 
1997-1998, 2002-2003 or 2004-2005. However, 2006-2007 outburst (middle 
panels) shows a delay pattern between 5.0-12.0 keV and 1.5-3.0 energy bands.

The two DCF display a remarkable similarity in the case of XTE J1550-564 (lower 
right panel) during its 1998 outburst. It is also to be noted that the outburst 
duration of XTE J1550-564 is the lowest as compared to the other two outbursts. 
\subsection{Short-time-scale}
\label{qpo_power_lag}
We analysed X-ray transient XTE J1650-500 using the public archival 
observations from {\it RXTE} during the 2001 outburst using 
HEASARC\footnote{\url{http://heasarc.gsfc.nasa.gov/}} and restricted our 
study to observations when QPOs were detected at low frequency ($0.1-10$ Hz).

\subsubsection{{\bf QPO and Fractional {\it {\bf power}}}}
\label{power}

\begin{figure}
\begin{center}
\includegraphics[height=0.75\columnwidth,width=1.0\columnwidth]{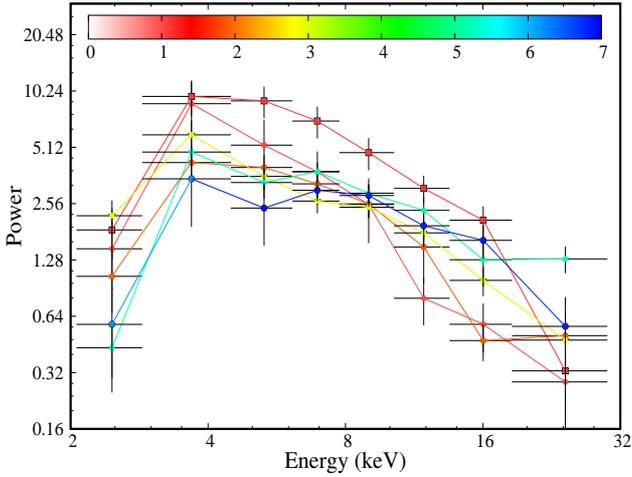}
\caption{Variations of power of QPOs with respect to the energy are presented 
in log scale. Colorbar represents the QPO centroid frequency. The maximum power 
for lower frequency QPOs are seen within $3-5$ keV energy range and 
for higher $\nu_c$ (> 3 Hz), the peak of the {\it power} is almost 
flat up to 8 keV.
\label{fig3}}
\end{center}
\end{figure}
Using the 
{\tt GHATS}\footnote{\url{http://astrosat.iucaa.in/~astrosat/GHATS_Package/Home.html}} 
software, we analysed Good Xenon, Event, and Single Bit data modes which 
contain high time-resolution data for timing analysis. In channel band 0-35 
(2-15 keV), we generated PDSs of every 16s for each observation. For each 
observation, we combined them to get an average PDS and subtracted the 
Poissonian noise contribution \cite{Zh95}. The PDSs are standardized and 
converted into squared fractional rms {which we used as `Power'}.
The power spectra were then fitted with Lorentzian (\ref{fig2}) 
combinations \citep{no00}, using {\tt XSPEC} version 12.0.

Complex behaviour of power density spectra in  XTE J1650-500 (see, \ref{fig2})
has already been reported in \cite{Ho03}.
The power density spectra consist of a primary frequency associated with a higher 
harmonic. Typically, the primary becomes stronger with increasing energy 
range. In the case of XTE J1650-500, the PDS remains enigmatic 
as compared to any other outbursting Galactic black hole candidates. 
Several sub-harmonics and higher harmonics present along with the primary,
are reported in \cite{Ho03}. In Fig. \ref{fig2}, we observed the need 
for an absorption model such as {\tt gabs} near 1 Hz during fitting 
where the signature of {\tt gabs} is more prominent in the lower energy range. 

The {\it power} variations over various energy bins are presented in Fig. \ref{fig3}. 
We used EVENT mode data sets where the channel groups used in current studies are 
the following: 0-6 (2.0-2.87 keV), 7-10 (2.87-4.49 keV), 11-14 (4.49-6.12 keV), 
15-18 (6.12-7.76 keV), 19-24 (7.76-10.22 keV), 25-32 (10.22-13.52 keV), 33-44 
(13.52-18.52 keV), 45-70 (18.52-29.97 keV). 

The {\it power} associated with QPOs is maximum at around $\sim 4 $ keV for lower QPO
frequency $\nu_c$ < 3 Hz. The peak of the {\it power} is almost flat up to $\sim 8$ keV 
for $\nu_c$ > 3 Hz and starts to decrease after that. Contrary to this, 
GS 1124-68, GX 339-4 \citep{Be97}, XTE J1550-564 \citep{CZCM99}, 
GRS 1915+105 \citep{cm00} exhibited more power in the higher energy (> 9 keV) range. From the colorbar of 
Fig. \ref{fig3}, we find that the QPO centroid frequency ($\nu_c$) remains 
almost constant with respect to the energy bin in each observation. Apart from the 
peak position of {\it power}, the weakening of the QPO {\it power} with respect to
the $\nu_c$ is similar to what is noticed in other outbursts.

\subsubsection{{\bf Time lags associated with QPOs }}
\label{s_qpo_lag}

We calculated the cross-spectrum which is defined as, 
$CF\left(j\right) = X^{\ast}_{1}\left(j\right) X_{2}\left(j\right)$,
where $X_{1}$ and $X_{2}$ are the complex Fourier coefficients for
the two energy bands at a frequency $\nu_{j}$ and $X^{\ast}_1\left(j\right)$
is the complex conjugate of $ X_{1}\left(j\right)$ (van der Klis et al. 1987).
The phase lag between the signals of two different energy bands at
Fourier frequency $\nu_{j}$ is, $\phi_j$=$arg\left[CF\left(j\right)\right]$
(i.e., $\phi_j$ is the position angle of $CF\left(j\right)$ in the complex plane).
The corresponding time lag is $\Delta t_j$ = $\frac{\phi_j}{2\pi\nu}$.
An average cross vector $C$ is determined by averaging the complex values
for every stretch of time. In our analysis, we produced a time 
lag spectrum for each observation and divided the data into two
energy bands, namely, soft (2-5 keV) and hard (5-13 keV). We extracted cross-spectra 
from 16s intervals, which are then averaged yielding one time-lag spectrum 
for each observation. Positive time-lag indicates that the harder photons lag
the softer photons. Following \cite{Re00}, we calculated the 
time lags at the QPOs to obtain a better understanding of the photons emitted 
from the Compton cloud (please see \cite{cm00}).

\begin{figure}
\begin{center}
\includegraphics[width=1.0\columnwidth,angle=0]{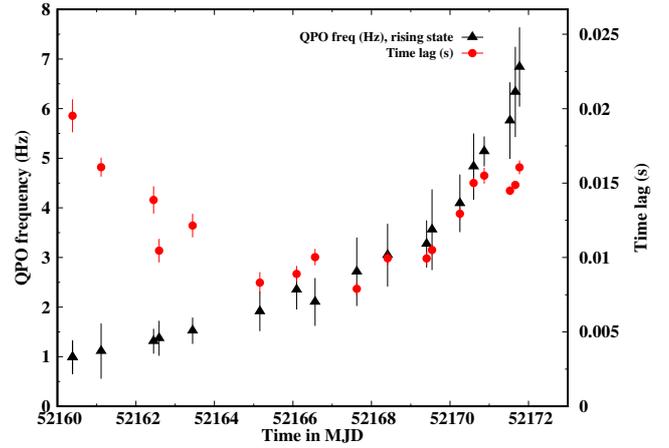}
\caption{Evolution of QPO centroid frequency $\nu_c$ (black-triangle) during 
the rising phase of the outburst. Associated time lags are plotted in red-circle.
\label{qpo_lag}} 
\end{center}
\end{figure}

QPO frequency evolution (black-triangle in Fig. \ref{qpo_lag}) during the hard 
and hard intermediate state is smooth and monotonically increasing as seen 
in other outbursts. A similar type of QPO evolution can be seen in case of 
almost every outbursting Galactic black holes and are observed for GX 339-4 
and XTE J1550-564, and the feature is generic irrespective of their 
inclination angle. We fitted the {\it RXTE}/{\tt pca} spectrum using 
{\tt wabs*(diskbb + powerlaw + gaussian)} where the photon index ($\Gamma$)
varied from 1.6-2.0 during the rising phase. We find $\nu_c$ and $\Gamma$
correlates (using 18 observations during the rising phase) with each other having 
Pearson correlation coefficient = 0.897, Spearman's rank ($\rho$) = 0.896, 
and p-value < .001.

\begin{table}
\centering

    \caption{Time lag Energy Correlation}     
    \label{tab:correlation}

    \begin{small}
    \begin{tabular}{|c|c|c|c|c|}
    \hline
{\bfseries Sources} & {\bfseries No. of Points} & {\bfseries PCC} & {\bfseries $\rho$}  & {\bfseries p-value } \\
    \hline
\textcolor{black}{XTE J1650-500}  &  20   & 0.743              &   0.727  & .000175    \\
    \hline
\textcolor{black}{GX 339-4}       &  20   & 0.681              &   0.777  & .000948    \\
    \hline
\textcolor{black}{XTE J1550-564}  &  20   & 0.197              &   0.156  & .407        \\
    \hline
    \end{tabular}
    \end{small} 
\end{table}

The time lags associated with QPOs (red-circle in Fig. \ref{qpo_lag}) evolve during the 
rising phase. Up to MJD 52168, the time lag decreases as the QPO frequency 
increases and showed an anti-correlation between them. But, after MJD 52168, 
both QPOs and lags exhibit a steady increase in the magnitude until it reaches 
a saturation value ($\sim 0.015$ sec). 

\begin{figure*}
\vskip 1.5cm
\begin{center}
\includegraphics[height=0.6\columnwidth,width=2.0\columnwidth,angle=0]{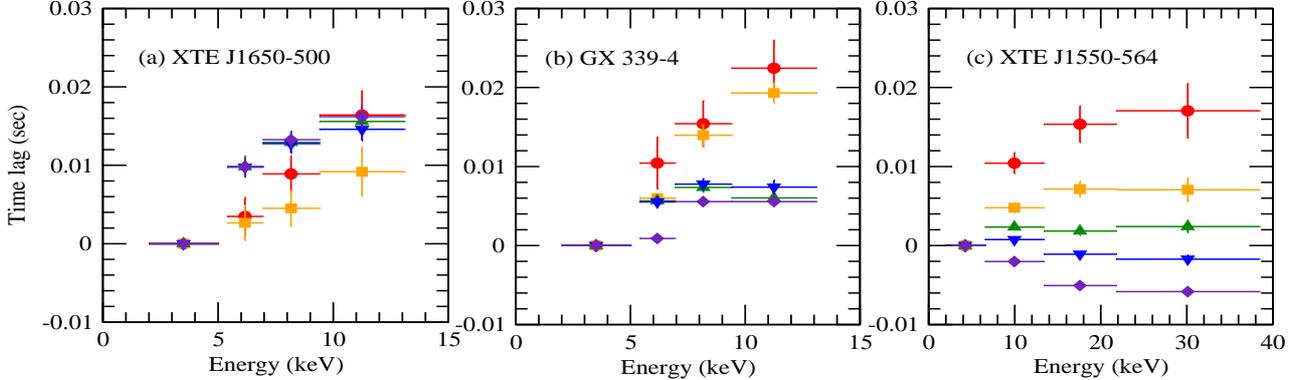}
\caption{Time lags with respect to energies are plotted. QPO centroid frequencies 
are marked with colour. Colour Red (circle), Orange (square), Green (triangle up), 
Blue (triangle down) and Violet (diamond) represent the increasing QPO frequencies 
of each sources. From panel (a), a strong correlation between time lag and 
energy can be observed for all QPO frequencies of XTE J1650-500, while similar correlations are seen in 
GX 339-4 for lower QPO frequencies (panel b). In panel (c), XTE J1550-564 exhibited energy dependent time 
lag patterns for all type of QPOs. But, the parameters do not correlate with each other. See Table 
\ref{tab:correlation} for parameter values. Panel b and c are adopted from Dutta \& Chakrabarti (2016).
\label{lag_energy}}
\end{center}
\end{figure*}

The time lag variation XTE J1650-500 with energy is presented in panel (a) of Fig. \ref{lag_energy}. 
Lag spectra are plotted for various QPO centroid frequencies. The hard lag  
increases with the increase of the energy irrespective of $\nu_c$ and have the 
Pearson correlation coefficient (PCC) = 0.743, Spearman's rank $\rho=0.727$, and p-value 
< .001. Similar correlation is observed for GX 339-4 (PCC = 0.681, $\rho$ = 0.777, 
p-value < .001), but, has not been observed for XTE J1550-564 
where PCC = 0.197, $\rho$ = 0.156, p-value = .407 (see panel (b) and (c) 
of Fig. \ref{lag_energy}). Note that the lags characteristics in higher 
($\nu_c$ > 3 Hz) QPO frequencies Fig. \ref{lag_energy}a do not change much.

In Fig. \ref{lag_comp}, we plot this time lag evolution along with those for 
GX 339-4 (blue) during its 2006-2007 outburst and XTE J1550-564 (red) during 
its 1998 outburst. GX 339-4 and XTE J1550-564 have reported inclinations 
of around $\sim 50^{\circ}$ and $\sim 70^{\circ}$ respectively. According 
to \cite{Ro05}, XTE J1650-500 has an approximate inclination of 
$\sim 45^{\circ}$. It is clear that the lag evolution of this
source followed a separate trail than the other two.

\subsubsection{\bf Time lags associated with iron line}
\label{iron}

\begin{figure}
\begin{center}
\includegraphics[height=0.7\columnwidth,width=1.0\columnwidth,angle=0]{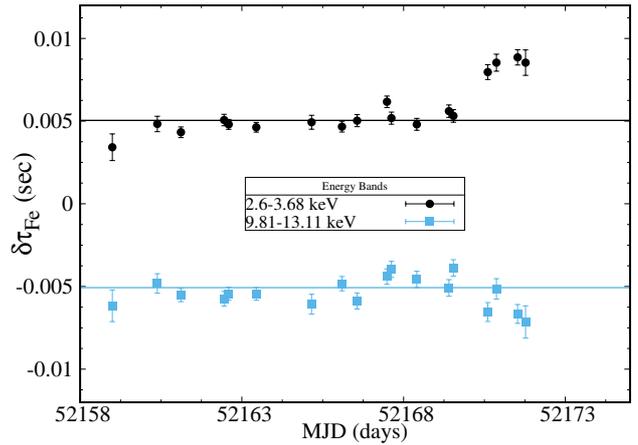}
\caption{Progress of lag of iron line photons with respect to soft 
(black) and Comptonized (cyan) photons. The fitted mean lags are 
represented by the solid lines.
\label{iron_lag}}
\end{center}
\end{figure}

\cite{Ro05}, found an anti-correlation between the Fe-line flux and the 
hard X-ray flux during the hard state. They reported a near-constant Fe-line 
flux which is almost $10\%-20\%$ of the power-law flux in the hard state. 
The reversal of the correlation line between Fe-line flux and power-law 
flux found in their work strongly suggests the presence of a region where 
the ionization state of the disc might have changed. Considering the 
contribution of iron line flux with respect to the power-law 
flux, we studied the evolution of the iron line lags (in Fig. 
\ref{iron_lag}). During the hard state, we examined the lags of iron line 
with respect to the soft (2.6-3.68 keV, say $\delta \tau_{disc}$) 
and hard photons (i.e., Comptonized photons, 9.81-13.11 keV, say $\delta \tau_{comp}$). 
We considered RXTE {\tt pca} 
channel 0-8 for the soft and 23-31 channels for Compton up-scattered 
photons. For the sake of uniformity, we considered channels 13-16, 
i.e., 5.7-6.94 keV for the Fe-line photons. The line-flux and equivalent 
width (200-500 eV) remains almost constant throughout the rising phase 
and declines in the intermediate state \citep{Ro05}.
To obtain the lag, we integrated the lag spectrum considering the centroid located 
at 5 Hz with FWHM of 5 Hz, which covers the $0.1-10$ Hz LFQPO domain. This range also 
represents the region of high coherence in the time lag spectrum.

From Fig. \ref{iron_lag}, we find a roughly constant lag with respect to the soft 
and Comptonized photons up to MJD 52168. During this period, the source evolved  
substantially, and the QPO frequency went up from 0.98 Hz to $\sim 3$ Hz.
After MJD 52168, the absolute magnitude of lags due to the iron line
with respect to the soft and Compton up-scattered photons 
increases. Thus, the Fe-line photon lags with respect to the soft photon 
(black-circle) and leads with respect to the Comptonized photon (cyan-square) 
during the hard to intermediate states. In essence, $\tau_{comp} > 
\tau_{Fe} > \tau_{soft}$ where $\tau_i$s represent the absolute time 
of arrival of Comptonized, Fe-line and soft photons respectively. 
The consistency of the Fe-line lag with respect to the soft 
and Comptonized photon indicates that the region generating  
the iron line emission remained similar up to
MJD 52168 and changes afterwards. \cite{Re13}
found the reflection fraction ($R$) less than unity during MJD 52158 to
MJD 52168. Later, $R$ increased sharply until the soft intermediate state.
After that, the variation of $R$ became erratic. It was suggested that 
the spectra were not light bending dominated during the rising phase.

\section{Discussions}
\label{discussion}
In this work, we observed the evolution of time lags associated with the QPOs, the
long time ($12\pm1$ days) scale delay of hard photons with respect to the soft component,
the {\it power} associated with QPOs is maximum around lower energy range ($\sim 4 - 8$ keV)
and constant lags of broad Fe-line photons with respect to soft and comptonized photons. We have also
shown comparative studies with some other outbursts such as GX 339-4 and XTE J1550-564.
Below, we present possible explanations of the timing variabilities which makes this 
outburst unique.   

\subsection{$12\pm1$ days delay}
\label{viscous_delay}
We found a significant delay of softer photons (1.5-3.0 keV) by 
$12\pm1$ days (Fig. \ref{fig1}) with respect to harder 
component (5.0-12.0 keV), though no significant 
delay was found between 1.5-3.0 keV and 
3.0-5.0 keV bands (top right panel Fig. \ref{fig1}) using three techniques 
of cross-correlation to confirm the delay. Analogous results are              
observed for GX 339-4 (middle panel of Fig. \ref{fig1}), where even larger 
($16\pm1.1$ days) delay was observed between the similar
energy bands during its 2006-2007 outburst. However, we did not find 
any delay patterns in the previous outbursts of GX 339-4. The zero-delay
patterns are also seen in XTE J1550-564 during its 1998 outbursts where
both of the {\tt DCF} patterns coincide with each other. 

In Two Component Advective Flow (TCAF) paradigm, this $12 \pm 1$ days delay
could be interpreted to be due to the viscous delay of the Keplerian component, i.e. disc component
\citep{CT95, GC19} with respect to the hot advective halo component 
(i.e., lower angular momentum component) which reaches
the inner region with almost the free-fall velocity. The Keplerian matter 
moves inward at viscous time-scale and enhances the cooling rate 
resulting in the shrinking of the post-shock region of the advective halo behaving here
as the Compton cloud. This causes the delay of softer photons. In the case of 
XTE J1550-564, the overlap suggests that the disc size could be small and the two 
components of accretion flow moved in simultaneously 
during its 1998 outburst, much like Cyg X-1 as reported in \cite{sm02}.

\subsection{QPO evolution}
\label{reducing_cloud}
From Fig. \ref{qpo_lag}, we see a steady increase in QPO frequency (black-triangle)
during the rising phase of the outburst. The fitted curve (Fig. \ref{qpo_lag}) shows 
that $\nu_c$ evolves as 
$\sim t^{1.72}$. According to Two Component Advective Flow model, the size of the Compton 
cloud is directly related to the QPO frequency (see \cite{cm00}) through a 
relation, $\nu_c=\left(c/r_g\right)/\left[Rr_s(r_s-1)^{1/2}\right]\sim r_s^{-3/2}$, 
where $c$ is the velocity of light, $r_g$ is the Schwarzschild radius, 
$r_s$ is the location of the centrifugal pressure supported shock of 
compression ratio $R$. Thus, the Compton cloud size varies roughly as 
$r_s \sim \nu_c^{-2/3} \sim (t^{1.72})^{-2/3} \sim t^{-1.14}$, 
indicating a sharp reduction of Compton cloud compared to XTE J1550-564 
(see, \cite{CDP09} for further details). \cite{Mo09} associated the drifting 
of power spectra towards the higher frequency regime in terms of the reduction 
in the radius of the Compton cloud. Similar results were obtained by \cite{YW12}, 
where the reduction of size was found to be a factor of $\sim$23 times during 
the spectral evolution. They concluded that the separation of mass accretion 
rate and Compton cloud are the critical drivers for spectral evolution which was 
predicted earlier by \cite{CT95} where two accretion rates in two components 
are assumed to be the cause of the evolution of the QPOs and spectra.
The halo rate controls the size and thermodynamical properties of Compton cloud, and 
disc rate causes the cooling of the Compton cloud resulting in the reduction of its size.
Thus, the evolution of QPO frequency can be linked to the reducing size of the Compton cloud 
and is generic among the majority of the outbursting sources irrespective of their inclination
angle.

\subsection{Disc-Jet Connection in XTE J1650-500}
\label{disc-jet}
Evidence of disc-jet connections from the timing properties is observed 
in various sources \citep{Ga08, AM15, Ve17, Re18}. Through simulations, 
\cite{RK19} showed a correlation between time lag-$\Gamma$ using the jet 
model. Recently, a suggestion was made that the soft lags \citep{Ch19} 
of XTE J1550-564 during its outburst in 1998 could be jet induced. 
Following that, \cite{pa19} inspected four high inclination GBHs where 
the presence of soft lags are seen during higher activity in radio 
fluxes. During the rising phase of XTE J1650-500, \cite{Co04} reported 
radio activity and \cite{Cu12} analysed optical-NIR wavelengths where 
pieces of evidence of X-ray photons reprocessed within the jet medium were
observed. Here, we explain a few observational results based on
disc-jet connections.

\subsubsection{Stronger QPOs at lower energy:}
\label{power}
QPO frequency can be directly linked to the size of the Compton cloud while
the energy-dependent QPOs and time lags associated with the QPOs capture the 
information on thermodynamic fluctuations present within the Compton cloud (see \cite{Ch17b}). 
We performed energy-dependent QPO {\it power} analysis (see Fig \ref{fig3}), 
where we find the {\it power} maxima of low-frequency QPOs reside within 4.0-8.0 
keV energy range. The steeper peak was observed for lower frequencies (< 3 Hz) 
while the higher frequency QPOs (> 3 Hz) exhibited flatter peaks. 

\cite{Be97} observed that the higher energy photons (9.3-37.2 keV) are participating 
more in the QPO formation of GS 1124-68 and GX 339-4 during their hard state. 
\cite{CZCM99} observed similar behaviour in the energy-dependent QPOs of XTE J1550-564.
\cite{cm00} found that the power maxima of GRS 1915+105 in the range of 
$\sim 9.0-13.0$ keV, which led them to conjecture that the Compton cloud alone 
takes part in oscillations. Later, this was also established by works
of \cite{vad01, rao00, ch05}. The energy-dependent {\it power} spectrum
of XTE J1650-500, however, shows a peak at $\sim 4.0-8.0$ keV range, just above
the disc temperature.

Considering the possibility of a low inclination source (see \cite{Sa02}, \cite{Ro05}),
there exists a likelihood where a fraction of emergent Comptonized
photons would interact with the electrons in the outflow. During the first
few days, when the source is in the hard state, size of the Compton cloud, 
which serves as the base of the jet, is profuse (see \cite{Ch99}). The
jets are compact and could down-scatter the emergent photons leading to a
shift of the peak of the power towards the lower energy as shown in
Fig. \ref{fig3} for QPOs having $\nu_c$ < 3 Hz.
As the source reaches to intermediate states, the jet becomes stronger
and denser (see \cite{Ja17}) where the possibility of up-scattering along
with down-scattering rises. We conclude that this could be a reason for
flatter peak in the fractional power with frequencies higher than $\sim$ 3
Hz.

\subsubsection{Time lag evolution:}
\label{lag_evolution}
The PCC and $\rho$ between time lag and energy is 0.743 and 0.727 respectively 
(see \S \ref{s_qpo_lag} and Table \ref{tab:correlation} for details) which suggest 
that the time lags are significantly dependent on the energy. According to \cite{p80}, 
Compton delay would yield larger time lags for harder photons and the lag is directly 
related to the size of the Compton cloud. Therefore, the current observation implicates 
that Comptonization is the dominating mechanism for the generation of lag during the 
rising phase as was also observed in the case of GX 339-4. The correlation dramatically 
drops (< 0.2) for the high inclination source XTE J1550-564 indicating 
the importance of other physical mechanisms, such as, reflection, gravitational 
bending, feedback from the outflows in the resultant lag evolution.

\begin{figure}[h]
\begin{center}
\includegraphics[width=1.0\columnwidth,angle=0]{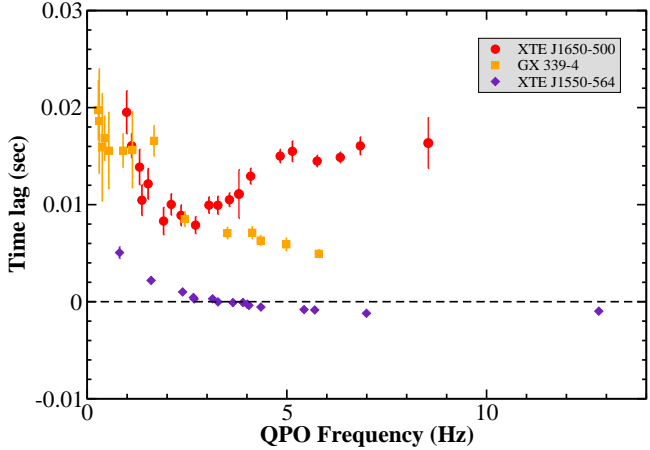}
\caption{A comparative diagram of time lag and QPO frequency correlation 
is drawn for three sources during the rising phases of the outbursts. 
Each of the sources follows a particular trend. Interestingly, GX 339-4 
and XTE J1550-564 followed similar patterns where the time lag decreased 
with increasing QPO frequency. On the contrary, the time lag initially 
decreased for XTE J1650-500 up to $\nu_c\sim 3.0$ Hz and afterwards 
increased with increasing QPO frequency.
\label{lag_comp}}
\end{center}
\end{figure}

From Fig. \ref{qpo_lag}, we see a systematic decrease in the time lags as 
QPO centroid frequency increases. However, the pattern reverses for XTE J1550-564 
after MJD 52168 and the time lag starts to increase. In TCAF paradigm, the 
post-shock region (Compton cloud or CENBOL) Comptonizes the soft photons to 
produce hard photons, and LFQPOs are produced due to the oscillations of the Compton cloud. 
The QPO frequency, in such cases, is inversely related to the size of the 
Compton cloud. Thus, the increment in QPO frequency is always expected
when the Compton cloud's size is reduced as the object evolves to a softer
state. The shrinking in Compton cloud reduces the delay due to Compton scattering (see \cite{Mi88}).
This is what we observe in the case of GX 339-4 and XTE J1550-564 (see Fig. \ref{lag_comp}).
Monte-Carlo simulations \citep{Ch17b} in the presence of Comptonization \citep{PSS83}, 
disc reflection and gravitational bending of photons \citep{Ch17a, Ch18} 
showed a direct anti-correlation with the size of the Compton cloud as seen 
in case of GX 339-4.

In the present object, however, the behaviour was similar when QPO frequency 
was less than 3 Hz, i.e., when the Compton cloud size is large, and the shock 
was strong with the hot post-shock region. Jets and outflows produced
from Compton cloud surface would be driven away from the system which would not contribute
to any further up-scattering. As the source evolves towards the intermediate 
state, the amount of outflow increases (as stated above in \S \ref{power}), 
and becomes dense and compact. Thus, the photons originated from Compton
cloud could have up-scattered in the sub-sonic region of the jet before reaching to the observer.
Therefore, the time lag increases gradually instead of getting reduced with increasing
frequency, when the source undergoes the transition from hard to the intermediate
state. The time lag saturates with respect to QPO frequency after reaching to
the intermediate state. \cite{RK19} simulated time lag-photon index ($\Gamma$) 
correlation using the jet model where the electron velocity is chosen in the outward 
direction. Increase of $\Gamma$ indicates spectral softening, and in general 
correlates with the $\nu_c$ (for XTE J1650-500, PCC > 0.85) during rising 
and declining phase. Thus, the correlation between time lag-$\nu_c$ after 
MJD 52168 can be explained by considering the jet medium. However, \cite{RK19} calculated the
lags in the frequency domain 0.05-5.0 Hz, not around the QPO centroid frequency.
Thus, the anti-correlation between time lag-$\nu_c$, which is observed for numerous 
black hole candidates, is absent in their work.

\subsubsection{Iron line lag:}
\label{jet_iron}
As evident from the \S \ref{iron}, iron line photons (5.7-6.94 keV) lag with respect to 
soft photons (2.6-3.68 keV) and lead with respect to comptonized photons (9.81-13.11 keV) 
during the rising state. A consistent lag magnitude with respect to both soft 
and Comptonized photons can be observed up to MJD 52168. 

\cite{Re13} showed the variation of reflection fraction ($R$) and photon index with days
(Fig. 3 and 8 respectively) where the $R$ less than one was observed during
the initial rising phase and started to rise after $\sim$ MJD 52168 where we 
observed the magnitude of lag corresponding to iron line started to 
increase with respect to disc (i.e., soft) and Comptonized photons. 
Considering the arrival delay of the Keplerian disc, as seen in Fig. \ref{fig1} and 
a relatively lower value of the reflected fraction, the light bending dominated 
reflection mechanism (see \cite{MF04}) could be less presiding. 

On the other hand, if we consider activities in radio wave (see \cite{Co04}), 
it remains a possibility that the Compton scattering in the jet 
medium (see \citet{CT95} and 
\citet{LT07} for details) might have produced the iron line 
during the hard and hard-intermediate states. In the absence of a Keplerian
disc, the accretion was dominated by optically thin halo during the rising 
phase. In that case, the receding jet could have also contributed to broadening 
of the iron line. 

The variations of QPO {\it power}, time lags associated with the QPOs, 
and lags of iron line with respect to soft and hard photons during the rising 
phase favours the low inclination scenario (<35\textdegree) of 
XTE J1650-500 (as suggested by \cite{Sa02}) in which case the interception 
of Comptonized photons in the jet region could maximize before reaching 
the observer.

\section{Conclusions}
\label{conclusion}

\begin{enumerate}
\item[1.] A positive delay of $12\pm1$ days between 1.5-3.0 keV to 
5.0-12.0 keV energy range is found from the ASM data. Various correlation 
algorithms corroborate this. Although, no definite delay can 
be ascertained between the energy bands 1.5-3.0 keV to 3.0-5.0 keV band. From 
a theoretical point of view, this delay of soft photons with respect to 
their harder counterpart could be a result of viscous delay due to 
the presence of a large Keplerian disc.\\ 

\item[2.] The {\it power} variation of QPOs with energy are atypical compared to 
other Galactic black holes. For lower frequency QPOs (i.e., $\nu_c < 3$ Hz), the 
{\it power} maxima are found just above the so-called disc or thermal part of the spectrum 
(around $\sim 4$ keV). But, for QPOs with $\nu_c > 3$ Hz, the {\it power}-energy 
curve became flatter.\\

\item[3.] The time lags between iron to disc (i.e., soft) and iron to Comptonized 
photons remain almost constant up to MJD 52168. The absolute value 
of both the lag increases afterwards as the outburst progresses towards 
the intermediate state. \\ 

\item[4.] The time lag evolution pattern follows a unique path 
where initially the lag magnitude decreases with increasing QPO 
frequency, i.e., anti-correlates up to $\nu_c \sim 3$ Hz. Later 
on, after MJD 52168, the lag magnitude is correlated 
with the QPO frequency. \\

Since Comptonization governed the lag-energy 
spectra in the entire rising phase of XTE J1650-500, higher lag at higher frequency 
implies a secondary source of Comptonization becoming dominant since 
the original Compton cloud was shrinking. We claim that this 
secondary source is the outflow originated from the Compton cloud. 
This is possible only if the inclination angle is low. Thus,
the variability analysis and comparative studies of time lags lead us to
conclude that the outflow or jet could play a major role in controlling the 
temporal properties of XTE J1650-500 during its outburst in 2001.
\end{enumerate}

\section*{Acknowledgements}
We acknowledge the anonymous Reviewer for helpful suggestions
which improved the clarity of the manuscript. We thank T. Belloni 
for providing the timing analysis software {\tt GHATS}.
AC acknowledges postdoctoral fellowship of S. N. Bose National Centre 
for Basic Sciences under the Department of Science and Technology (DST), Govt. 
of India. BGD acknowledges IUCAA for the Visiting Associateship 
Programme. PN acknowledges CSIR fellowship for this work.
This research has made use of data and/or software provided by the 
High Energy Astrophysics Science Archive Research Center (HEASARC), 
which is a service of the Astrophysics Science Division at NASA/GSFC 
and the High Energy Astrophysics Division of the Smithsonian
Astrophysical Observatory.

\section*{Data Availability}
The data underlying this article are available in the HEASARC
archive at \url{https://heasarc.gsfc.nasa.gov}.




\bibliography{reference}

\begin{thebibliography}{}
\bibitem[{Alexzander}(1997)]{A97}Alexander T. 1997, ASSL, 218, 163
\bibitem[{Altamirano \& M\'endez}(2015)]{AM15}Altamirano D. \& M\'endez M., 2015, MNRAS, 449, 4027
\bibitem[{Belloni \& Hasinger}(1990)]{BH90}Belloni T., \& Hasinger G., 1990b, A\&A, 230, 103
\bibitem[{Belloni et al.}(1997)]{Be97}Belloni T., van der Klis M., Lewin W. H. G., van Paradijs J., Dotani T.,Mitsuda K. \& Miyamoto S., 1997, A\&A, 322, 857
\bibitem[{B\"ottcher \& Liang}(1999)]{BL99}B\"ottcher M. \& Liang E. P., 1999, ApJ, 511, L37
\bibitem[{Cabanac et al.}(2010)]{c10}Cabanac C., Henri G., Petrucci P.-O., Malzac J., Ferreira J. \& Belloni T. M., 2010, MNRAS, 404, 738
\bibitem[{Corbel et al.(2003)}]{Co03}Corbel S., Nowak M. A., Fender R. P., Tzioumis A. K., Markoff S., 2003, A\&A, 400, 1007
\bibitem[{Chakrabarti \& Titarchuk}(1995)]{CT95}Chakrabarti S. K., Titarchuk L. G., 1995, ApJ, 455, 623
\bibitem[{Chakrabarti}(1999)]{Ch99}Chakrabarti S. K., 1999, A\&A, 351, 185
\bibitem[{Chakrabarti \& Manickam}(2000)]{cm00}Chakrabarti S. K. \& Manickam S., 2000, ApJ, 531, 41
\bibitem[{Chakrabarti et al.}(2005)]{ch05}Chakrabarti S. K., Nandi A., Chatterjee A. K., Choudhury A. K. \& Chatterjee U., 2005, A\&A, 431, 825
\bibitem[{Chakrabarti et al.}(2009)]{CDP09}Chakrabarti S. K., Dutta B. G., \& Pal P. S. 2009, MNRAS, 394, 1463C
\bibitem[{Chatterjee, Chakrabarti \& Ghosh}(2017a)]{Ch17a}Chatterjee A., Chakrabarti S. K. \& Ghosh H., 2017, MNRAS, 465, 3902
\bibitem[{Chatterjee et al.}(2017b)]{Ch17b}Chatterjee A., Chakrabarti S. K. \& Ghosh H., 2017, MNRAS, 472, 1842
\bibitem[{Chatterjee et al.}(2018)]{Ch18}Chatterjee A., Chakrabarti S. K., Ghosh H. \& Garain S., 2018, MNRAS, 478, 3356
\bibitem[{Chatterjee et al.}(2019)]{Ch19}Chatterjee A., Dutta B. G., Patra P., Chakrabarti S. K. \& Nandi P., 2019, Proceedings, 17, 8, doi:10.3390/proceedings2019017008
\bibitem[{Corbel et al.}(2004)]{Co04}Corbel S., Fender R. P., Tomsick J. A., Tzioumis A. K., \& Tingay S. 2004, ApJ, 617, 1272
\bibitem[{Cui, Chen \& Zhang}(1999)]{CCZ99}Cui W., Chen W., \& Zhang S. N., 1999, ApJ, 484, 383
\bibitem[{Cui et al.}(1999)]{CZCM99}Cui W., Zhang S. N., Chen W., \& Morgan E. H., 1999, ApJ, 512, L43
\bibitem[{Curran et al.}(2012)]{Cu12}Curran P. A., Chaty S. \& Zurita Heras J. A., 2012, A\&A, 547, A41
\bibitem[{Done \& Gierli\`nski}(2006)]{DG06}Done C. \& Gierli\`nski M., 2006, MNRAS, 367, 659
\bibitem[{Dutta \& Chakrabarti}(2016)]{dc16}Dutta B. G. \& Chakrabarti S. K., 2016, ApJ, 828, 101
\bibitem[{Dutta, Pal \& Chakrabarti}(2018)]{dpc18}Dutta B. G., Pal P. S. \& Chakrabarti S. K., 2018, MNRAS, 479, 2183
\bibitem[{Edelson \& Krolik}(1988)]{ek88}Edelson R. A. \& Krolik J. H., 1988, ApJ, 333, 646
\bibitem[{Fabian \& Vaughan}(2003)]{FV03}Fabian A. C. \& Vaughan S., 2003, MNRAS, 340, L28
\bibitem[{Gandhi et al.}(2008)]{Ga08}Gandhi P. et al., 2008, MNRAS, 390, L29-L33	
\bibitem[{Garain et al.}(2014)]{ga14}Garain S. K., Ghosh H. \& Chakrabarti S. K., 2014, 437, 1329 
\bibitem[{Gaskell \& Peterson}(1987)]{ga87}Gaskell C. M. \& Peterson B. M., 1987, ApJS, 65, 1
\bibitem[{Ghosh \& Chakrabarti}(2019)]{GC19}Ghosh \& Chakrabarti, 2019, MNRAS, 484, 5802
\bibitem[{Heil et al.}(2015)]{he15}Heil L. M., Uttley P. \& Klein-Wolt M., 2015, MNRAS, 448, 3348
\bibitem[{Homan et al.}(2003)]{Ho03}Homan J., Klein-Wolt M., Rossi S., et al. 2003, ApJ, 586, 1262
\bibitem[{Jana et al.}(2017)]{Ja17}Jana A., Chakrabarti S. K., Debnath D., 2017, ApJ, 850, 91
\bibitem[{Laurent \& Titarchuk}(2007)]{LT07}Laurent P. \& Titarchuk, L., 2007, ApJ, 656, 1056
\bibitem[{Lightman \& Eardley}(1974)]{LE74}Lightman A. P. \& Eardley D. M., 1974, ApJ, 187, 1
\bibitem[{Lin et al. }(2000)]{l00}Lin D., Smith I. A., Liang E. P. \& B\"ottcher M., 2000, ApJ, 543, L141
\bibitem[{Ingram et al.}(2009)]{in09}Ingram A., Done C., Fragile P. C., 2009, MNRAS, 397, L101
\bibitem[{Ingram \& van der Klis}(2015)]{iv15}Ingram A. \& van der Klis M., 2015, MNRAS, 446, 3516
\bibitem[{Kalamkar et al.}(2015)]{Ka15}Kalamkar M., van der Klis M., Heil L. \& Homan J., 2015, ApJ, 808, 144
\bibitem[{Kalemci et al.}(2003)]{Ka03}Kalemci E., Tomsick J. A., Rothschild R. E., et al. 2003, ApJ, 586, 419
\bibitem[{McClintock \& Remillard}(2006)]{MR06}McClintock J. E., \& Remillard R. A. 2006, in Compact Stellar X-ray Sources, eds. W. Lewin, \& M. van der Klis (Cambridge: Cambridge Univ. Press), 157
\bibitem[{Merloni et al.}(2003)]{Me03}Merloni A., Heinz S., di Matteo T., 2003, MNRAS, 345, 1057
\bibitem[{Miyamoto et al.}(1988)]{Mi88}Miyamoto S., Kitamoto S., Mitsuda K., \& Dotani T. 1988, Natur, 336, 450
\bibitem[{Miller et al.}(2002)]{Mi02}Miller J. M., Fabian A. C., Wijnands R., et al. 2002, ApJ, 570, L69
\bibitem[{Miniutti \& Fabian}(2004)]{MF04}Miniutti G., \& Fabian A. C. 2004, MNRAS, 349, 1435
\bibitem[{Miniutti et al.}(2004)]{Mi04}Miniutti G., Fabian A. C., \& Miller J. M. 2004, MNRAS, 351,466
\bibitem[{Molteni et al.}(1994)]{Mo94}Molteni D., Lanzafame G. \& Chakrabarti S. K., 1994, ApJ, 425, 161
\bibitem[{Montanari et al.}(2009)]{Mo09}Montanari E., Titarchuk L., \& Frontera F. 2009, ApJ, 692, 1597
\bibitem[{Motta et al.}(2015)]{Mo15}Motta S. E., Casella P., Henze M., Mu\~noz-Darias T., Sanna A., Fender R. \& Belloni T., 2015, MNRAS, 447, 2059
\bibitem[{Nowak et al.}(1999)]{NWD99}Nowak M. A., Wilms J., Dove J. B., 1999, ApJ, 517, 355
\bibitem[{Nowak}(2000)]{no00}Nowak M. A., 2000, MNRAS, 318, 361
\bibitem[{Orosz et al.}(2004)]{Or04}Orosz J. A., McClintock J. E., Remillard R. A., \& Corbel, S. 2004, ApJ, 616, 376
\bibitem[{Patra et al.}(2019)]{pa19}Patra D., Chatterjee A., Dutta B. G., Chakrabarti S. K. \& Nandi P., 2019, ApJ, 886, 137
\bibitem[{Payne}(1980)]{p80}Payne D. G., 1980, ApJ, 237, 951
\bibitem[{Poutanen \& Fabian}(1999)]{PF99}Poutanen J., Fabian A. C. 1999, MNRAS, 306, L31
\bibitem[{Pozdnyakov, Sobol \& Sunyaev}(1983)]{PSS83}Pozdnyakov A., Sobol I. M., Sunyaev R. A., 1983, Astrophys. Space Sci. Rev., 2, 189
\bibitem[{Rao et al.}(2000)]{rao00}Rao A. R., Naik S., Vadawale S. V. \& Chakrabarti S. K., 2000, A\&A, 360, L25
\bibitem[{Reig et al.}(2000)]{Re00}Reig P., Belloni T., van der Klis M., M\'endez M., Kylafis N. D. \& Ford, E. C, 2000, ApJ, 541, 883
\bibitem[{Reig et al.}(2018)]{Re18}Reig P., Kylafis N. D., Papadakis I. E., \& Costado M. T. 2018, MNRAS, 473, 4644
\bibitem[{Reig \& Kylafis}(2019)]{RK19}Reig P. \& Kylafis N. D., 2019, A\&A, 625, 90
\bibitem[{Reis et al.}(2013)]{Re13}Reis R. C., Miller J. M., Reynolds M. T. , Fabian A. C., Walton D. J. et al., 2013, ApJ, 763, 38
\bibitem[{Remillard}(2001)]{Re01}Remillard R. 2001, IAU Circ., 7707, 1
\bibitem[{Rossi et al.}(2004)]{Ro04}Rossi S., Miller J. M., Homan J., \& Belloni T. 2004, Memorie della Societa Astronomica Italiana Supplementi, 5, 184
\bibitem[{Rossi et al.}(2005)]{Ro05}Rossi S., Homan J., Miller J. M., \& Belloni T. 2005, MNRAS, 360, 763
\bibitem[{Rutledge et al.}(1999)]{Ru99}Rutledge R. E. et al., 1999, ApJS, 124, 265
\bibitem[{Sanchez-Fernandez et al.}(2002)]{Sa02}Sanchez-Fernandez C., Zurita C., Casares J., Castro-Tirado A. J., Bond, I., Brandt, S., \& Lund, N. 2002, IAU Circ., 7989, 1
\bibitem[{Schnittman et al. }(2006)]{s06}Schnittman J. D., Homan J. \& Miller J. M., 2006, ApJ, 642, 420
\bibitem[{Shakura \& Sunyaev}(1973)]{SS73}Shakura N. I., \& Sunyaev R. A. 1973, A\&A, 24, 337	
\bibitem[{Smith et al.}(2002)]{sm02}Smith D. M., Heindl W. A., Markwardt C. B. \& Swank J. H., 2001, ApJ, 554, L41
\bibitem[{Stella \& Vietri}(1999)]{SV99a}Stella L.,\& Vietri M., 1999, Phys.Rev.Lett.,82,17
\bibitem[{Sunyaev \& Titarchuk}(1980)]{ST80}Sunyaev R. A. \& Titarchuk L. G., 1980, A\&A, 86, 121
\bibitem[{Tagger \& Pellat}(1999)]{tp99}Tagger M. \& Pellat R., 1999, A\&A, 349, 1003
\bibitem[{Tetarenko et al.,}(2019)]{T19}Tetarenko A. J., Sivakoff G. R., Miller-Jones J. C. A., Bremer M. \& Mooley K. P. et al., 2019, MNRAS, 482, 2950
\bibitem[{Tomsick et al. }(2004)]{To04}Tomsick J. A., Kalemci E., \& Kaaret P. 2004, ApJ, 601, 439
\bibitem[{Vadawale et al.}(2001)]{vad01}Vadawale S. V., Rao A. R. \& Chakrabarti S. K., 2001, A\&A, 372, 793
\bibitem[{van den Eijnden et al.}(2017)]{van17}van den Eijnden J., Ingram A., Uttley P., Motta S. E., Belloni T. M., Gardenier D. W., 2017, MNRAS, 464, 2643
\bibitem[{van der Klis et al.}(1987)]{van87}van der Klis M., Hasinger G., Stella L., et al. 1987, ApJ, 319, L13
\bibitem[{van der Klis}(2004)]{van04}van der Klis M., arXiv:astro-ph/0410551
\bibitem[{Veledina et al.}(2017)]{Ve17}Veledina A. et al., 2017, MNRAS, 470, 48
\bibitem[{Yan \& Wang}(2012)]{YW12}Yan L. H., \& Wang J. C. 2012, RAA, 12, 269
\bibitem[{Zhang et al.}(1995)]{Zh95}Zhang W., Jahoda K., Swank J. H., Morgan E. H. \& Giles A. B., 1995, ApJ, 449, 930
\end{thebibliography}
{}
\bsp	
\label{lastpage}
\end{document}